\documentclass[aps,prb,amsmath,twocolumn,amssymb,titlepage,superscriptaddress,showpacs]{revtex4-1}

\usepackage{epsfig}
\usepackage{graphicx}
\usepackage{lipsum}
\usepackage{dcolumn}
\usepackage{bm}
\usepackage[utf8]{inputenc}
\usepackage{graphicx} 
\usepackage{xcolor}
\usepackage{float} 
\usepackage[normalem]{ulem}

\newcommand{\I}{\mathrm{i}}

\newcommand{\ave}[1]{\langle #1 \rangle}

\newcommand{\green}[2]{\langle \mathcal{T_C} #1(t) #2(t') \rangle}
\newcommand{\beq}[1]{\begin{equation} #1 \end{equation}}
\newcommand{\bsplit}[1]{\begin{equation} \begin{split} #1 \end{split} \end{equation}}

\newcommand{\astcycl}{\mathrlap{\kern0.085em{\circlearrowright}}\ast}
\newcommand{\taucycl}{\mathrlap{\kern0.42em{\bullet}}\circlearrowright}

\begin{document}

\title{Photo-induced phase transition and associated time scales in the excitonic insulator Ta$_2$NiSe$_5$}

\author{Tanusree Saha }
\altaffiliation{Corresponding authors: tanusree.saha@student.ung.si, primoz.rebernik@elettra.eu }
\affiliation{Laboratory of Quantum Optics, University of Nova Gorica, 5001 Nova Gorica, Slovenia.}

\author{Denis Gole\v{z} }
\affiliation{Center for Computational Quantum Physics, Flatiron Institute, 162 5th Avenue, New York, NY 10010, USA}
\affiliation{Department of Theoretical Physics, Institute Jo\v{z}ef Stefan, Jamova 39, SI-1001 Ljubljana, Slovenia. } 
\affiliation{Department of Complex Matter, Institute Jo\v{z}ef Stefan, Jamova 39, SI-1001 Ljubljana, Slovenia. } 

\author{Giovanni De Ninno}
\affiliation{Laboratory of Quantum Optics, University of Nova Gorica, 5001 Nova Gorica, Slovenia.}
\affiliation{Elettra-Sincrotrone Trieste, Area Science Park, 34149 Trieste, Italy.}

\author{Jernej Mravlje }
\affiliation{Department of Theoretical Physics, Institute Jo\v{z}ef Stefan, Jamova 39, SI-1001 Ljubljana, Slovenia. } 

\author{Yuta Murakami}
\affiliation{Department of Physics, Tokyo Institute of Technology, Meguro, Tokyo 152-8551, Japan}

\author{Barbara Ressel}
\affiliation{Laboratory of Quantum Optics, University of Nova Gorica, 5001 Nova Gorica, Slovenia.}

\author{Matija Stupar}
\affiliation{Laboratory of Quantum Optics, University of Nova Gorica, 5001 Nova Gorica, Slovenia.}

\author{Primo\v{z} Rebernik Ribi\v{c} $^\ast$}
\affiliation{Laboratory of Quantum Optics, University of Nova Gorica, 5001 Nova Gorica, Slovenia.}
\affiliation{Elettra-Sincrotrone Trieste, Area Science Park, 34149 Trieste, Italy.}

\begin{abstract}
We investigate the non-equilibrium electronic structure and characteristic time scales in a candidate excitonic insulator, Ta$_2$NiSe$_5$, using time- and angle-resolved photoemission spectroscopy with a temporal resolution of 50 fs. Following a strong photoexcitation, the band gap closes transiently within 100 fs, i.e., on a time scale faster than the typical lattice vibrational period. Furthermore, we find that the characteristic time associated with the rise of the photoemission intensity above the Fermi energy decreases with increasing excitation strength, while the relaxation time of the electron population towards equilibrium shows an opposite behaviour. We argue that these experimental observations can be consistently explained by an excitonic origin of the band gap in the material. The excitonic picture is supported by microscopic calculations based on the non-equilibrium Green's function formalism for an interacting two-band system. We interpret the speedup of the rise time with fluence in terms of an enhanced scattering probability between photo-excited electrons and excitons, leading to an initially faster decay of the order parameter. We show that the inclusion of electron-phonon coupling at a semi-classical level changes only the quantitative aspects of the proposed dynamics, while the qualitative features remain the same. The experimental observations and microscopic calculations allow us to develop a simple and intuitive phenomenological model that captures the main dynamics after photoexcitation in Ta$_2$NiSe$_5$.
\end{abstract}
\date{\today}

\maketitle
\section{Introduction}

Excitonic insulators were originally proposed as materials where a spontaneous condensation of excitons takes place in the vicinity of the semiconductor-semimetal transition\cite{mott1961,knox1963}. Such a condensation is associated to the violation of separate charge conservation in the conduction and valence bands~(internal $U(1)$ gauge symmetry). While this phase has been unambiguously identified for heterostructures in the quantized Hall regime~\cite{butov2002,eisenstein2004bose,wang2019evidence}, only a few single-material systems have been suggested to exhibit a ground-state excitonic insulator phase. Among them, Ta$_2$NiSe$_5$\cite{wakisaka2009excitonic,wakisaka2012} and 1T-TiSe$_2$\cite{cercellier2007} are the most prominent ones.  However, in real materials, there are always effects that may break the $U(1)$ symmetry\cite{mazza2020}. In addition, the presence of electron-phonon coupling~\cite{kaneko2013, murakami2017,subedi2020_prm}  can lead to a structurally ordered phase with properties similar to that of an excitonic insulator. Indeed, Ta$_2$NiSe$_5$ shows a structural phase transition from an orthorhombic to a monoclinic lattice at $328$K \cite{kaneko2013,disalvo1986}.  This raises an important long-standing question: what is the dominant origin of the order in Ta$_2$NiSe$_5$ ? A number of experiments have been performed, employing different methods such as Raman spectroscopy~\cite{kim2020direct,kim2020_prr,volkov2020critical}, optical techniques~\cite{larkin2017,larkin2018}, and angle-resolved photoemission spectroscopy (ARPES)~\cite{seki2014excitonic,watson2020_prr}, to answer this question. However, the results were interpreted in either the excitonic or phononic picture and no consensus has been reached so far.

Tracking the system as it evolves through a nonequilibirum phase transition triggered by a short optical pulse can provide further clues on the origin of the symmetry breaking in Ta$_2$NiSe$_5$. However, also in this case, the situation is controversial, as several groups performing time-resolved ARPES (trARPES) experiments have reported rather contradicting results \cite{mor2017,okazaki2018,baldini2020,tang2020,suzuki2020detecting}. Some studies~\cite{mor2017,baldini2020} found a  short-lived increase of the gap, which was interpreted as a transient enhancement of the excitonic order~\cite{mor2017}, stimulating further theoretical studies~\cite{murakami2017,tanaka2018,tanabe2018,ryo2019}. On the other hand, several recent experimental works have observed a decrease of the gap~\cite{mor2017,okazaki2018,tang2020}, albeit marginal in recent high-resolution data~\cite{baldini2020}, that is naturally expected as the excited electrons can scatter with condensed excitons~\cite{golez2016}.  Differences have also been observed in the measured characteristic time scales of the photoemission signal after photoexcitaion. Ref.~\onlinecite{okazaki2018} reports a characteristic time scale of around 100 fs, which is faster than the typical lattice oscillation in Ta$_2$NiSe$_5$.  On the other hand, the authors of Ref.~\onlinecite{baldini2020} argue that this time scale is comparable to the lattice distortion period (characteristic time above 250 fs) and consider this as an evidence of a lattice-driven transition. More generally, the behaviour of the time-scales vs. the excitation fluence is also controversial. In particular, this concerns the apparent disagreement between the dynamical slowing down, observed in both experimental \cite{zong2019_prl, tomeljak2009} as well as some theoretical studies focused on symmetry-broken states ~\cite{tsuji2013,tsuji2014,babadi2015,golez2016,babadi2017,picano2020,dolgirev2020}, and the measured rise time of the photoemission intensity above the Fermi energy, which becomes shorter for increasing photo-excitation strengths \cite{okazaki2018}. Summing up the above contrasting results, the question on the dominant driving force of the order in Ta$_2$NiSe$_5$ still remains open.

In this paper, we study the non-equilibrium electron dynamics of Ta$_2$NiSe$_5$ in a high photoexcitation regime, which was not explored in previous works. Using trARPES with a temporal resolution of 50 fs, we show that a strong photoexcitation induces a phase transition from an insulating to a semi-metallic phase and that the melting of the band gap occurs on a time scale faster than the period of a phonon oscillation. Next, we show that the dependence of different characteristic time scales on the photo-excitation strength shows contrasting behaviours. On one hand, the rise time of the photo-emission intensity above the Fermi energy becomes shorter with increasing excitation strength. On the other hand, the relaxation time of the electron population towards equilibrium becomes longer with increasing fluence, reaching saturation above a critical value.

We corroborate these experimental findings by carrying out numerical simulations accounting for the non-equilibrium response of a two-band tight-binding system, with interband Coulomb repulsion and an excitonic ground state. The calculations confirm closing of the band gap for high excitation strengths and the speed up of the photoemission signal in the conduction band with increasing fluence. They also allow us to analyse in detail the dynamics of the order parameter, which can be separated into two parts: a) a fast initial decay that speeds up with increasing fluence, b) a longer time scale that becomes slower on approaching the critical fluence, as shown in previous theoretical works~\cite{golevz2019dynamics,tsuji2013,babadi2015}. We provide a simple explanation for these different trends, showing their compatibility. In addition, we analyse the effect of electron-phonon coupling and show that it modifies only the quantitative features of these dynamics, while the qualitative aspects remain unchanged. Finally, we introduce a simple phenomenological model that captures the above dynamics and provides an intuitive picture of the physics after photoexcitation in an excitonic insulator. Our combined experimental/theoretical study supports a pre-eminently excitonic character of the order parameter in Ta$_2$NiSe$_5$.

\section{Experimental setup}

The dynamics of photo-excited electrons in Ta$_2$NiSe$_5$ was probed by means of trARPES. The measurements were performed at the CITIUS high-harmonic generation (HHG) light source~\cite{citius}. The system is driven by a mode-locked Ti:Sapphire laser delivering 800-nm pulses, with a duration of 40 fs at a repetition rate of 5 kHz. The driving laser was split in two beams: the major part of the intensity was used to generate extreme-ultraviolet (EUV) probe pulses through HHG, with Ar as the generating medium, and the remaining part was used as the pump. The intensity of the pump pulses on the sample was controlled with a variable attenuator - in all experimental plots, the fluence refers to the peak energy density (in mJ/cm$^2$), determined from the expression $2E_{p}/(\pi w^2)$, where $E_p$ is the  energy per pulse and $w$ is the beam waist at the sample position. Both pump and probe pulses were p-polarized, with an in-plane component parallel to the atomic chains of Ta$_2$NiSe$_5$ ($\mathbf{E}_{in} \parallel \mathbf{a}$) and an out-of-plane component perpendicular to the sample surface ($\mathbf{E}_{out} \parallel \mathbf{b}$). The photon energy of the probe was selected by a monochromator grating with off-plane geometry, which preserved the pulse duration~\cite{poletto}. Because Ni 3\textit{d} and Se 4\textit{p} electron orbitals have a high photo-ionisation cross-section at photon energies close to 26 eV, a probe energy of 26.35 eV (harmonic 17 of the seed laser) was used in the experiments. In order to preserve the ultrafast response, the energy resolution of the source was limited to about 150 meV. This allowed us to achieve a temporal resolution of around 50 fs.  The energy and momenta of the photo-emitted electrons were measured by a R3000 hemispherical analyser from VG Scienta. A closed-cycle Helium cryostat was used to control the sample temperature. The samples were cleaved in situ at a pressure lower than 6x10$^{-9}$ mbar to generate a clean mirror-like surface, prior to each measurement. The experiments were performed at a pressure lower than 1x10$^{-10}$ mbar.

\begin{figure*}[t]
\centering
\vspace{-4ex}
\includegraphics[width = 1\linewidth]{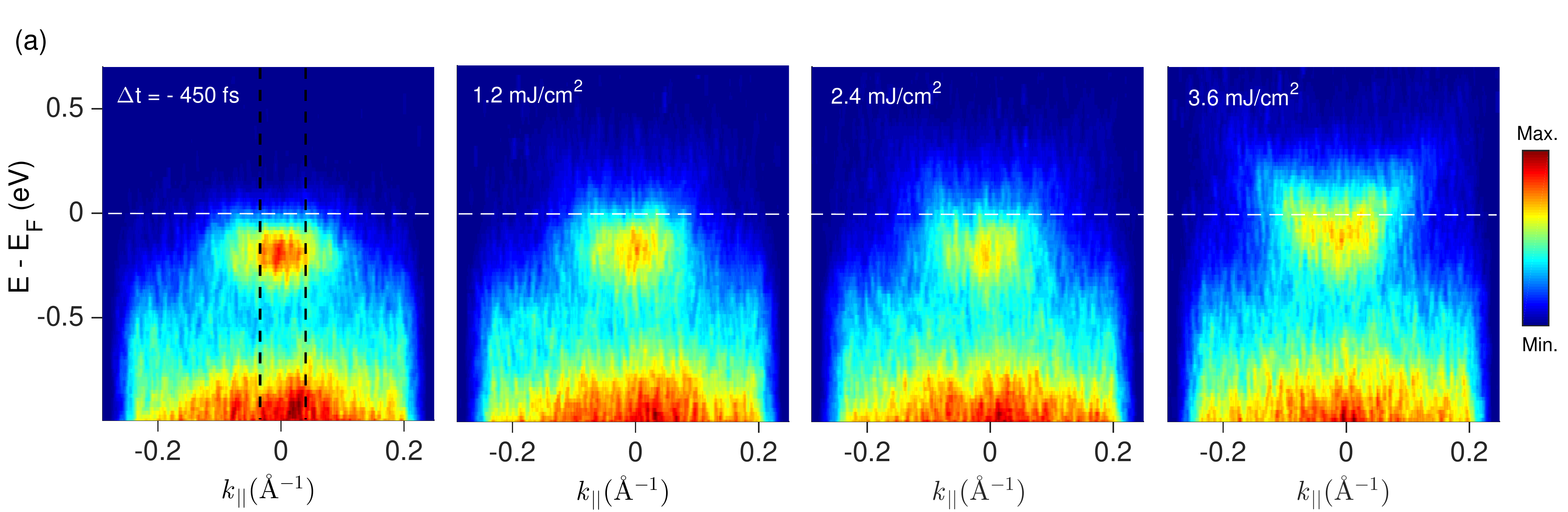}
\includegraphics[width = 1\linewidth]{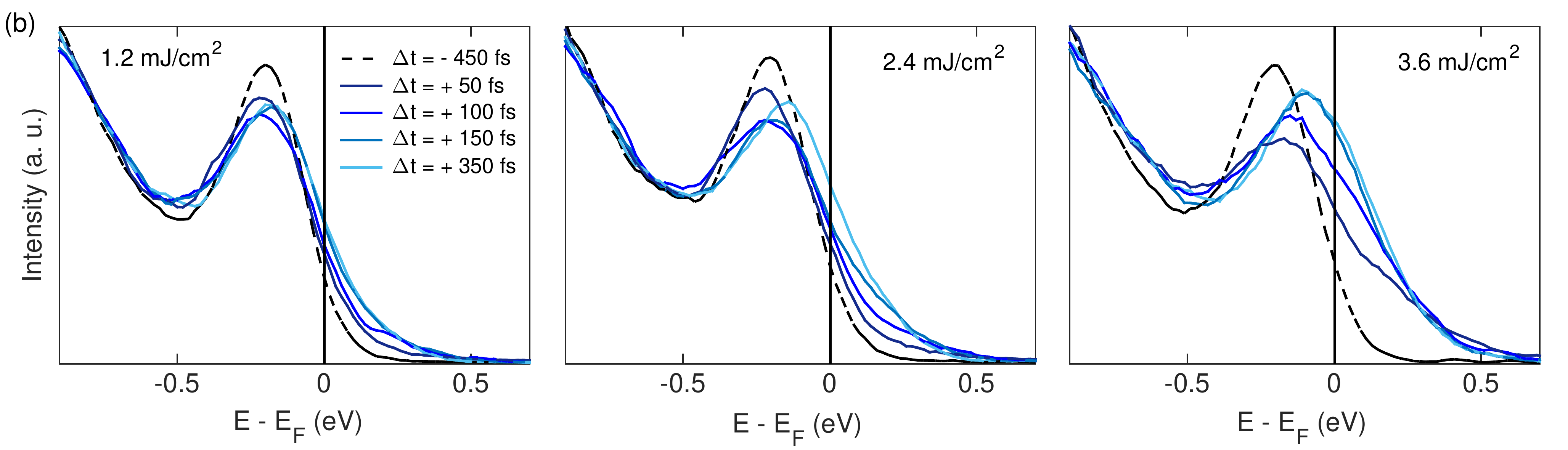}
\vspace{-2ex}
\caption{(a) Static (no pump) ARPES spectrum of Ta$_2$NiSe$_5$ at 100 K, measured using a photon energy of 26.35 eV (left-most panel), and trARPES snapshots at $\Delta t$ = 150 fs after photoexcitation for different pump fluences. The Fermi level is marked by the horizontal dashed line; (b) momentum-integrated EDCs at different delays ($\Delta t$), showing the temporal evolution of the gap dynamics for different pump fluences. The vertical dashed lines (at $\Delta t$ = -450 fs) indicate the considered region for momentum integration.}
\label{fig1}
\vspace{-2ex}
\end{figure*}

\section{Experimental results}

We investigated the evolution of the topmost valence band 
as a function of the incident pump fluence. This valence band is formed mainly by hybridized Ni 3$d$ and Se 4$p$ orbitals. As discussed in previous ARPES studies\cite{wakisaka2009excitonic,seki2014excitonic}, the band flatness is a possible indication of an excitonic phase in the material, which sets in below T$_c$ = 328 K.  The equilibrium (no pump) ARPES spectrum of Ta$_2$NiSe$_5$ around $\Gamma$ (Brillouin zone center)-point at a temperature of 100 K is
shown in Fig.~\ref{fig1}(a) (left-most panel). The Fermi level E$_F$, which is represented by the horizontal dashed line, is
determined with respect to the valence band spectrum of a clean gold
sample and is about 0.2 eV above the topmost valence band. After photoexcitation of the system with
an infrared pump pulse ($h\nu$ = 1.55 eV), the spectral weight of the
valence band decreases and is transferred to unoccupied energy levels above E$_F$. In Fig.~\ref{fig1}(a), one can observe that, at a pump fluence of 1.2 mJ/cm$^2$, the amount of spectral weight transfer from the topmost valence band to the conduction band is relatively small. Increasing the pump fluence results in an increase of the spectral weight transfer. While this increase is modest at  2.4 mJ/cm$^2$,  the spectral weight transfer is considerably enhanced for a pump fluence of 3.6 mJ/cm$^2$.

These observations are made more evident in Fig.~\ref{fig1}(b), where we show the momentum integrated energy distribution curves (EDCs) before the arrival of the pump pulse (at $\Delta$t = -450 fs) and at different delays after photoexcitation, as a function of fluence. The integrated angular range for all the EDCs is indicated by the black vertical lines in the
ARPES spectrum shown in the leftmost panel of Fig.~\ref{fig1}(a). Comparing the
EDCs before and after photoexcitation, we observe that for the lowest fluence of 1.2 mJ/cm$^2$, the dynamics of the valence band shows a depopulation and broadening, accompanied by a small energy shift towards the Fermi level at
subsequent delay times. At an intermediate pump fluence of 2.4 mJ/cm$^2$, which is above the value reported in previous studies, the spectral weight transfer to the conduction band is increased, while the energy shift of the valence band towards the Fermi level remains relatively small.  Finally, at a pump fluence of 3.6 mJ/cm$^2$, the energy shift of the valence band towards E$_F$ is pronounced and there is a significant increase of the photoemission intensity up to several hundreds of meV above the Fermi level. These observations suggest that a strong pump fluence can induce a phase transition in which the energy gap closes and the system is driven from an insulating to a semi-metallic state~\cite{okazaki2018photo,tang2020}.

The key information that can be extracted from the above results is the time scale after photoexcitation, within which the band gap closes. This provides details on the origin of the band gap \cite{hellmann2012time}, i.e., whether the dominant interaction governing the band dynamics is electronic or phononic. Figure 1(b) shows that, at a pump fluence of 3.6 mJ/cm$^2$, the band gap closes within $\sim$100 fs, a time interval that is shorter compared to the relevant vibrational time, which is about 500 fs for the 2~THz phonon mode)~\cite{mor2018,subedi2020_prm,kim2020direct,kim2020_prr,volkov2020critical}. Even faster response is observed at the highest pump fluence used in this study; see Fig.~\ref{fig3}. This observation suggests that the induced phase transition has a strong electronic character. Notice that the gap closing is seen only for the highest fluence in Fig.~\ref{fig1}, while for the two lower fluences, only marginal energy shifts are observed. This suggest the existence of a critical fluence and might also help explaining the relatively small energy shifts observed in some of the previous trARPES experiments\cite{mor2017,baldini2020}. We will discuss this issue further in the theoretical analysis.

\begin{figure*}[t]
\vspace{2ex}
\includegraphics[width = 1\linewidth]{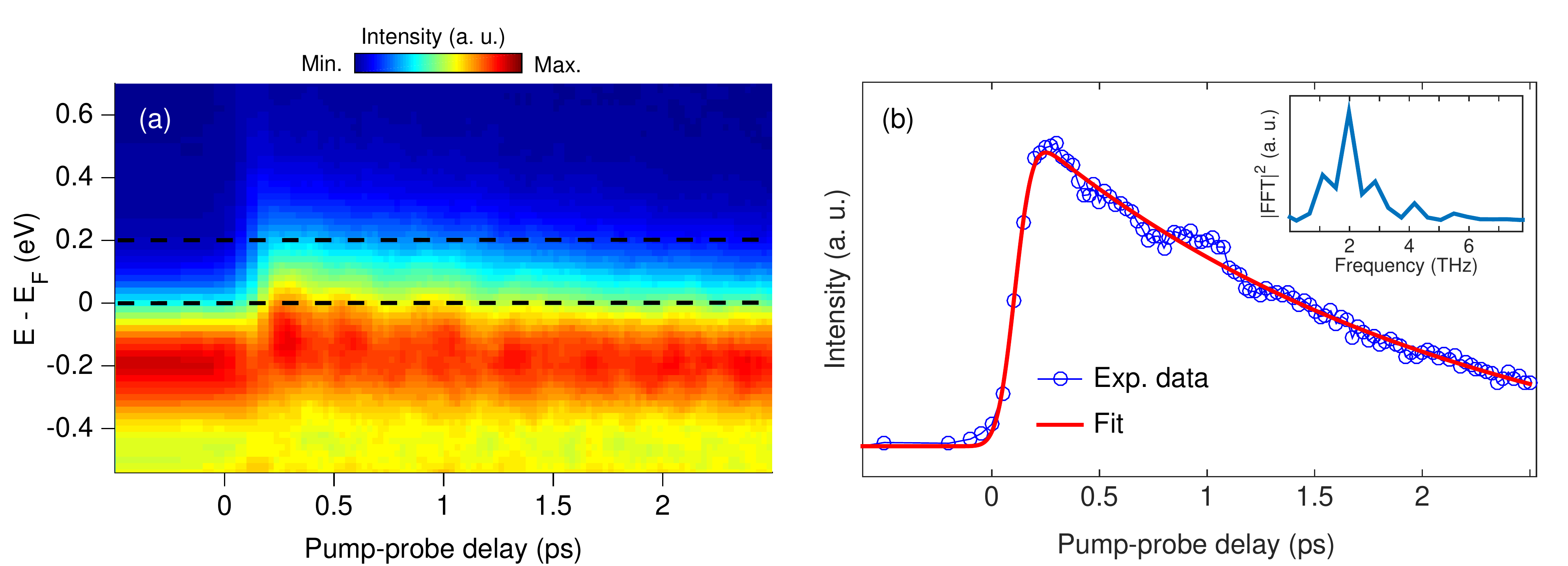}
\vspace{-2ex}
\caption{(a) trARPES spectrum around the $\Gamma$ point for the integrated angular range indicated by the vertical dashed lines in Fig.~\ref{fig1}(a). (b) Temporal evolution of the integrated intensity (filled blue circles) in the energy interval [0.0, 0.2] eV, shown by the horizontal dashed lines in (a). The red curve represents the fit (see text for details). The measurements were acquired at a pump fluence of 3.6 mJ/cm$^2 $.  Inset: Fourier analysis of the oscillatory part of the photoemission signal.}.
\label{fig2}
\vspace{-2ex}
\end{figure*}

We note here that the above findings are not fully conclusive. For example, one could argue that the relevant time scale for a pure structural phase transition is a quarter of the phononic oscillation only~\cite{ligges2018}, which is comparable to the observed time scale in Fig.~\ref{fig1}. In addition, the structural and electronic dynamics are coupled: the structural monoclinic displacement and the excitonic order act in a similar way on the electronic dispersion~\cite{mazza2020, subedi2020_prm}. Coherent oscillations in the photoemission spectrum with the relevant phonon frequency are therefore expected. Indeed, our data show a modulation of the integrated photoemission signal in the energy range from 0 to 0.2 eV, see Fig. 2b). However, the fact that the most intense phonon peak is located at 2 THz suggest that the lattice remains in its low-temperature phase even under relatively high fluences used in our experiment \cite{werdehausen2018, mor2018, Andrich2020phonon, kim2020direct}. This indicates that the dominant order is not  structural, as argued earlier in Ref.~\onlinecite{tang2020}. An excitonic origin of the band gap is further supported below, where we discuss the fluence dependence of the characteristic time scales observed in the photoemission signal.

Figure~\ref{fig2}(a) shows the angle-integrated photoemission spectral intensity as a function of electron binding energy and time delay
between the pump and probe pulses for a pump fluence of 3.6
mJ/cm$^2$. The temporal
evolution of the integrated intensity in the energy interval [0, 0.2]
eV is plotted in Fig.~\ref{fig2}(b). We observed two characteristic time scales: a fast rise time, $\tau_\mathrm{r}$, of the angle- and energy-integrated photoemission signal above Fermi energy and a slow relaxation time, $\tau_\mathrm{d}$, of the electron population towards equilibrium.  The intensity profile is 
fitted using the following function: \beq{ I(t)= H(t)
  [Ae^{(-t/\tau_d)}+B], } where \textit{H}(\textit{t}) describes the
finite rise-time of the intensity which is modelled by an error function, given by,
 \beq{ H(t)=1+\mathrm{erf}(t/\tau_c) ;}
  \textit{A} and $\tau_d$ are the amplitude and the decay time of the
  electronic response, respectively, and \textit{B} is a
  constant. Here, $\tau_c$ is the convolution of the natural
  electron-intensity rise time $\tau_r$ and the pump-probe
  cross-correlation time, which represents the finite temporal
  resolution of our system.
  
The energy-integrated, time-resolved photoemission intensity at different pump fluences is shown in Fig. 3(a). The rise time shows obvious shortening with increased excitation strengths, as emphasized by the dashed line marking the  peak position. In Fig. 3(b), we plot the rise time $\tau_r$, determined from fits shown in Fig. 3(a), as a function of pump fluence. The rise time has a maximum value of about 120 fs for the lowest pump fluence and decreases down to around 60 fs at the highest pump fluence used in the experiment. A fluence-dependent rise time further supports an excitonic origin of the order parameter in Ta$_2$NiSe$_5$ and will be interpreted in terms of an enhanced scattering probability between photo-excited electrons and excitons in the theoretical section. Although some degree of fluence dependence of the rise time might also occur in the purely phononic scenario, the phonon modes are expected to soften upon photoexcitation \cite{murray2007}, which would result in a trend opposite to the one observed in the experiment.

We also examined  the fluence dependence of the electron relaxation time $\tau_d$ towards equilibrium, see  Fig. 3(c). Contrary to the behaviour of  $\tau_r$, the relaxation time $\tau_d$ shows a monotonic increase of  up to the critical fluence, after which saturation is observed. This trend is consistent with previous experiments \cite{okazaki2018} and we attribute it to the scattering of phonons with excitons: at high excitation fluences a substantial population of excited vibrational modes can transfer energy back to the perturbed excitonic condensate and suppress its reestablishment \cite{Hedayat2019}.

\begin{figure*}
\vspace{-4ex}\includegraphics[width = 1\linewidth]{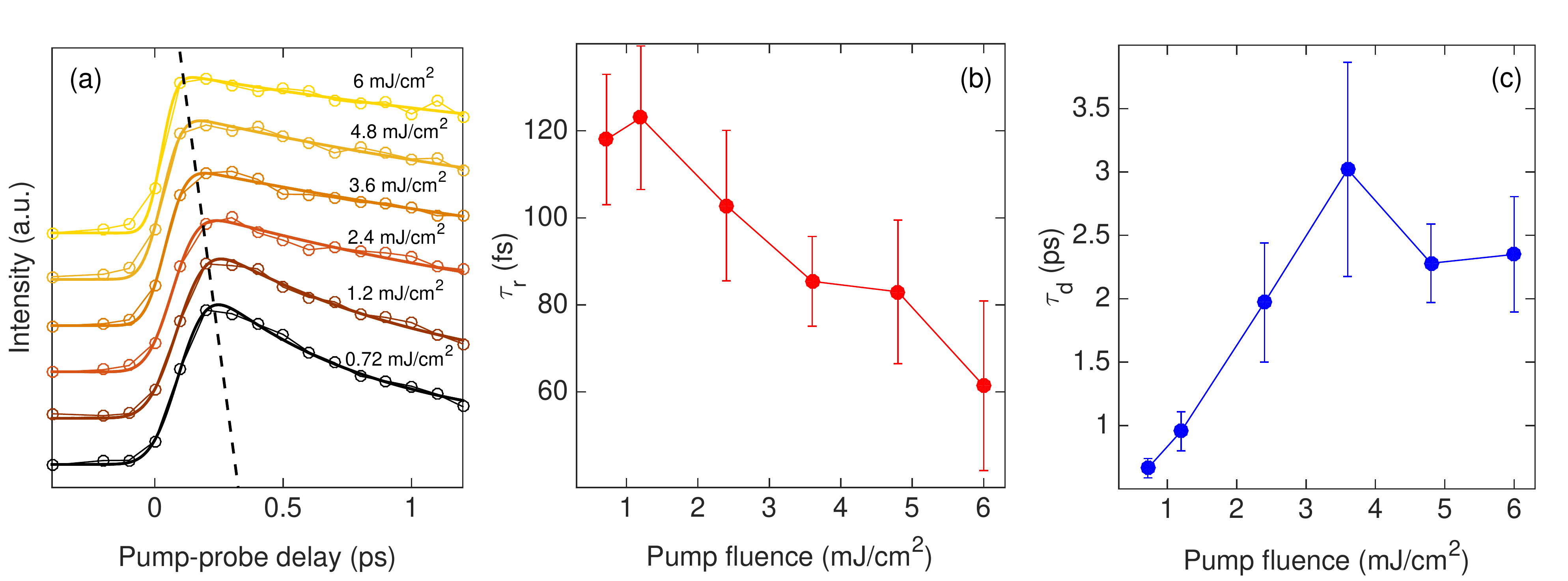}
\vspace{-2ex}
\caption{(a)  Energy-integrated photoemission intensity for several pump fluences, where the smooth curves are fits to the experimental data. The shortening of the rise time for increasing pump fluence is indicated by a black dashed line. (b) Rise time $\tau_\mathrm{r}$ as a function of pump fluence; (c) Dependence of the hot electron relaxation time $\tau_\mathrm{d}$ on pump fluence. 
}
\label{fig3}
\vspace{-2ex}
\end{figure*}

\section{Theoretical analysis}
\label{theory}

We now turn to the theoretical discussion, which will help us understand the early stages of the photoinduced dynamics in Ta$_2$NiSe$_5$. In the first subsection, we will introduce a model of an excitonic insulator. We will then discuss the theoretical results and compare them with the experimental findings shown in the previous sections. In the last subsection we will develop a simple phenomenological framework that captures the main dynamics after photoionization in an excitonic insulator.

\subsection{Model and method}

We consider a two-band tight-binding model in one dimension, which is appropriate for
Ta$_2$NiSe$_5$, due to the strong anisotropy along the
atomic chains~\cite{seki2011,seki2014excitonic}. For simplicity, we consider 
spinless fermions and couple the electrons in the two bands by a local interband repulsion. The total
Hamiltonian is given by
\bsplit{
	H=H_{\text{kin}}+H_{\text{int}}+H_{\text{dip}}+H_{\text{ph}}.
}
The kinetic energy reads
\begin{align}
H_{\text{kin}}=\sum_{\substack{k \\ \alpha \in \{0,1\}}} h_{k,\alpha} c_{k,\alpha}^\dagger c_{k,\alpha}
\label{Eq:Kin}
\end{align}
where $h_{k,\alpha}=-2J_\alpha\cos(k)-\mu+(-1)^{\alpha}\Delta\epsilon$ is the single-particle dispersion relation, $\Delta\epsilon$ the band-level splitting, and $J_0=-J_1$ the hopping integral; $c_{k,\alpha}^\dagger$ and $c_{k,\alpha}$ are, respectively, the creation and annihilation operators of an electron with momentum $k$ in the orbital $\alpha$. The chemical potential $\mu$ is chosen such that the system is at half filling. The interaction part includes the local interband density-density interaction with a strength $V$:
\bsplit{
H_{\text{int}}=V \sum_{i} n_{i,0} n_{i,1},
\label{Eq:Int}
}
where $n_{i,\alpha}=c_{i,\alpha}^{\dagger} c_{i,\alpha}$ is the electronic density at site $i$ and band $\alpha$. The Hamiltonian conserves the charge in individual bands, but at sufficiently low temperatures the
associated symmetry is spontaneously broken and the corresponding order parameter $\phi=\langle c_0^\dagger c_1\rangle$ acquires a finite value. In the previous expression, the operator index refers to the considered orbital, while the particle momentum is averaged. 
The order parameter $\phi$ is a complex quantity and represents the degree of spontaneous hybridization between the conducting and the valence bands (i.e., the number of condensed excitons).

We simulate the coupling of the electromagnetic field to the electrons by the term
\bsplit{
H_{\text{dip}}= E(t)  \sum_{i}   c_{i,1}^\dagger c_{i,0} + \text{h.c.},
\label{Eq:Dip}
}
where the index $i$ runs over all sites and $E(t)$ is the time-dependent amplitude of the electric field, which is responsible for the dipolar excitation between the conduction and valence bands.

In previous studies, the time evolution of the excitonic order was described within the time-dependent Hartree-Fock
theory~\cite{murakami2017,ryo2019}, which is however only applicable for
low photo-doping and short times, due to the lack of scattering and heating effects. In this work, we consider a strong photo-doping regime,
where the order is completely destroyed\cite{golez2016}. For this purpose, we employ the second Born approximation, describing the
non-equilibrium dynamics within the Keldysh formalism. This approach is non-perturbative in the strength of the electric field and properly describes electron-electron scattering, as well as the associated electron thermalization ~\cite{balzer2010,schlunzen2019,tuovinen2019,tuovinen2020_prb}.

The phonon term in the Hamiltonian is given by 

\bsplit{
  H_{\text{ph}}=g\sum_i (b_i+b_i^{\dagger}) (c_{i,1}^{\dagger}
  c_{i,0}+c_{i,0}^{\dagger} c_{i,1}) + \omega_0 \sum_{i} b_i^{\dagger}
  b_i.
\label{Eq:ph}} 
Here $b_i^{\dagger}$ and $b_i$ are, respectively, the phonon creation and annihilation operators and $g$ is the electron-phonon interaction strength. We consider the case of a dispersionless phonon with frequency $\omega_0$. We choose the form of the electron-phonon coupling in a way that is relevant for the mode involved in the orthorhombic to monoclinic
transition in Ta$_2$NiSe$_5$\cite{kaneko2013}, namely, a nonvanishing
phonon distortion leads to a hybridization of the two bands, and thus to a finite value of $\phi$. We treat the electron-phonon interaction on the mean-field level leading to a self-consistent classical equation of motion, see Appendix~\ref{App:2Born} for details.

\begin{figure*}[t]
  \includegraphics[width=1.0\linewidth]{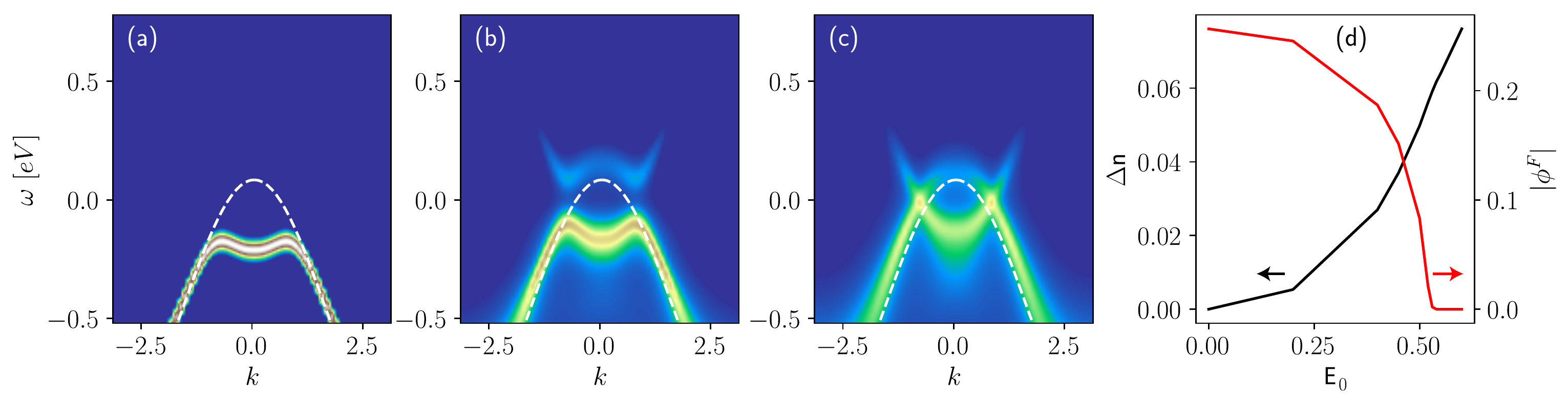}
  \caption{Snapshots of the simulated photoemission spectrum
  I$_k$($\omega$,t$_p$): a) at equilibrium; b) at a fixed pump-probe delay t$_p$=70 fs and $E_0=0.4$, i.e., below the value corresponding to the dynamical phase transition; c) at t$_p$=70 fs and $E_0=0.6$, above the critical value corresponding to gap closure. The Fermi level corresponds to $\omega=0$. The model parameters are $V/J_0=3.0$, $\lambda/J_0=0.0$ and the equilibrium state is prepared at $T=100$ K. The white dashed lines represent the valence band dispersion in equilibrium with an artificially suppressed order parameter. d) Dependence of the order parameter $|\phi^F|=|\phi(t\rightarrow\infty)|$ (red curve) and the polarization $\Delta n$ (black curve) on the excitation strength $E_0$ in the extrapolated long-time limit .\label{Fig:Akw}
}
\end{figure*}

The electron-phonon interaction includes two free parameters, namely the phonon frequency $\omega_0$ and the interaction strength $g$. We fix the phonon frequency to $\omega_0$=0.013 eV ~\cite{larkin2017,mor2018,werdehausen2018}. As both the electron-electron  $V$ and the electron-phonon $g$ interaction  can contribute to the gap opening, we have adjusted these two parameters so that they produce the same single-particle gap. At a mean-field level, the gap size due to the excitonic and lattice effects is governed by the effective interaction $V^*=V+2\lambda,$ where $\lambda=2g^2/(\omega_0)$. We have checked that the effective interaction $V^*$ is a good measure of the gap size in the framework of the second Born approximation and all subsequent calculations are done with fixed $V^*/J_0=3.0$. We also fixed the effective interaction $V^*$ and the crystal field $\Delta \epsilon$, such that the system is in the BCS parameter regime, where the normal~(disordered) phase is semimetallic.

\subsection{Theoretical results and comparison with experiment}
We have adjusted the theoretical calculations to the equilibrium ARPES spectrum and set the hopping parameter to $J_0=0.26eV$~\cite{mor2017},
which means that the inverse hopping corresponds to $\approx2.5$
fs. For easier comparison with the experimental part, we report
energies in units of eV and times in fs.
The system is excited with a short laser pulse that is parameterized as
\beq{ E(t)=E_0 \sin[\omega(t-t_0)]e^{-4.6(t-t_0)^2/t_0^2}, } where
$t_0=2\pi/\omega$ and $\hbar\omega$=1.5~eV, consistent with the experimental value. The pulse strength E$_0$ fixes the amount of photo-doping~(polarization). With these choices, the pulse duration is about $t\approx5$ fs. 

We start our analysis by simulating the time-dependent photoemission
spectrum, see Fig.~\ref{Fig:Akw}, which can be directly compared to the experimental data in Fig.~\ref{fig1}. Its expression is
$I_k(\omega,t_p)=-\text{Im}[\sum_{\alpha=0}^1\int dt' S^2(t')\exp^{\I \omega(t'-t)}G^<_{k,\alpha\alpha}(t',t),$ where $G^<_{k,\alpha\alpha}$ is the lesser component of the Green’s function (see Appendix A) and $S(t)=\exp(-t^2/2\delta^2)$ is the envelope of the probe pulse~\cite{freericks2009} with a duration $\delta=40$ fs, in agreement with the experimental setting.  At equilibrium, see Fig.~\ref{Fig:Akw}(a), only the states below the Fermi level are occupied and their dispersion shows a two-peak structure. This is characteristic of the so-called BCS regime, in which the excitonic correlation length is larger than the lattice spacing, and the corresponding disordered phase is semi-metallic. The excitonic hybridization between the conduction and valence bands opens up a gap at the degeneracy points~\cite{zenker2012,kaneko2013a}. After photoexcitation, the electrons are promoted from the valence to the conduction band. The valence band broadens and moves towards the Fermi level, see Fig.~\ref{Fig:Akw}(b). Above the critical value $E_0=0.51$, the gap is suppressed, see Fig.~\ref{Fig:Akw}(c), and one observes a semi-metallic dispersion, which looks consistent with the experimental results corresponding to the largest excitation strength in Fig.~\ref{fig1}. 

The reduction of the gap is driven by the suppression of excitonic
order. The value of the order parameter after a time much longer than the duration of the pump envelope is displayed in Fig.~\ref{Fig:Akw}(d). One observes a progressive suppression of the order parameter with fluence. Notice that only on approaching the critical fluence, the associated suppression of the gap becomes sizeable, which helps interpreting recent experiments~\cite{mor2017, baldini2020} that observed marginal effects on the spectra. Figure~\ref{Fig:Akw}(d) also shows the photo-induced polarization
\beq{
\Delta n(t)=\sum_{\alpha=0}^1\int_0^{\infty} d\omega
  [G^<_{\text{loc},\alpha\alpha}(\omega,t)-G^<_{\text{loc},\alpha\alpha}(\omega,0)]
 }
in the long-time limit. Increasing fluences leads to higher occupancy of the
conduction band, which corresponds to higher $\Delta n$ values.

\begin{figure*}
\includegraphics[width=0.8\linewidth]{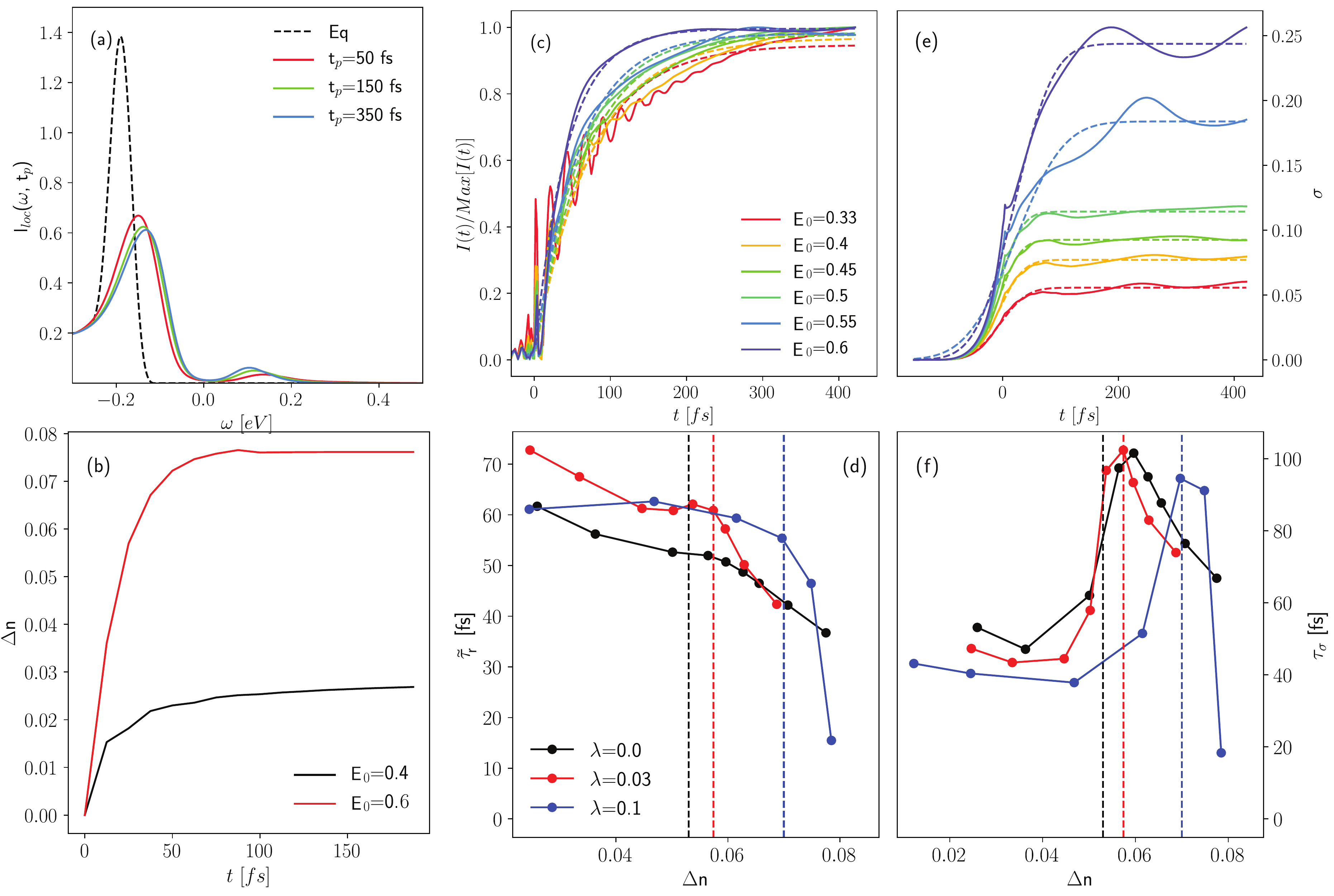}
\caption{
(a) Time evolution of the momentum averaged component of the photoemission spectra, after an excitation with a strength of $E_0=0.4$. The dashed line corresponds to the equilibrium spectrum (i.e., no pump). (b) Time evolution of the polarization $\Delta$n after excitation with $E_0=0.4$ and $E_0=0.6$. (c) Time evolution of the photoemission integral over the energy range~[0,~0.2] eV, for different excitation strengths.
(d) Rise time $\tau_I$ of the energy integrated photoemission signal versus the excitation strength, for the purely electronic theory~($\lambda/J_0=0.0$) and in the presence of a finite electron-phonon coupling ($\lambda/J_0=0.03$ and $\lambda/J_0=0.1$). The vertical lines mark the critical excitation strength, above which the order parameter, in the long-time limit, goes to zero. (e) Time evolution of the full-width at half maximum $\sigma$ for the dominant peak in the photoemission spectra (below the Fermi energy), at different excitation strengths. (f) Rise time $\tau_{\sigma}$ of the full-width at half maximum $\sigma$ for the purely electronic theory~($\lambda/J_0=0.0$) and in the presence of a finite electron-phonon coupling ($\lambda/J_0=0.03$ and $\lambda/J_0=0.1$).  
}
\label{Fig:tPES}
\vspace{-2ex}
\end{figure*}

We now turn to time dependence and start with the photoemission spectra. Figure~\ref{Fig:tPES}(a) shows snapshots of the momentum integrated photoemission spectrum $I_{\text{loc}}(\omega,t_p)=\frac{1}{N_k} \sum_k I_k(\omega,t_p)$, where $N_k$ (=128) is the number of momentum points we used in the simulation, at different delay times t$_p$. The data indicate a reduction of the gap size, and a substantial population of the conduction band. The evolution of the electronic population above the Fermi level shows a fast (order of 100 fs) relaxation. The number of photo-excited electrons, $\Delta n(t)$, increases even after the photoexcitation, Fig.~\ref{Fig:tPES}(b). We interpret this result by considering the scattering of electrons having excess kinetic energy with excitons. The resulting exciton dissociation generates two electrons and one hole. This leads to an increase in the number of conducting electrons, as it was previously observed in Ref.~\onlinecite{golez2016}. This anomalous increase in the number of conducting electrons was also observed in a different excitonic insulator candidate, namely 1T-TiSe$_2$, by analysing the growth of the photo-induced Drude weight in the time-resolved optical conductivity~\cite{mathias2016}.

For a direct comparison with the experiments reported in the previous section, we have integrated the momentum-integrated photoemission spectrum over the energy range [0-0.2] eV, as
$\langle I_{\text{loc}}\rangle(t_p)=\int_{0}^{0.2} d\omega
I_{\text{loc}}(\omega,t_p)$. In Fig.~\ref{Fig:tPES}(c), we plot  
$\langle I_{\text{loc}}\rangle(t_p)$, normalized to its maximal value, for several excitation strengths. For weak
excitation (i.e., $E_0\leq0.2$), $\langle
I_{\text{loc}}\rangle$ shows a small overall change and is rather
oscillating in time (data not shown), with a frequency corresponding to the amplitude mode excitation, i.e., $\approx$ 8 fs, well below our experimental resolution. For excitation strengths
closer to the critical value, the redistribution of the spectral weight is larger and the oscillations are suppressed.  Interestingly, at large fluences the dynamics of the intensity build-up shows a clear speed-up, consistent with experimental
findings.

In order to address this aspect more quantitatively, we fitted the curves shown in Fig.~\ref{Fig:tPES}(c) with a single exponential $I(t)=I_0 (1.0-\exp(-t/\tilde\tau_r))$, and extracted the rise time $\tilde\tau_r$. The latter is shown in Fig.~\ref{Fig:tPES}(d), as a function of polarization $\Delta n$, for different strengths of the electron-phonon coupling $\lambda$;  the vertical dashed lines indicate critical fluences. Let us first focus on the curve corresponding to $\lambda=0$, i.e., no electron-phonon coupling; we will turn to the other two cases later in the text. 

We note that both the magnitude of $\tilde\tau_r$ and the functional dependence with the excitation strength~(polarization) are in good agreement with measurements, see Fig. 3(b). The fluence dependence of the rise time is relatively weak for low fluences and becomes stronger at larger ones, showing a very weak trend in the region around the critical fluence. The observed discrepancy in the magnitude of experimental and  theoretical $\tau_r$ values can be explained by the strong
dependence of the latter on the microscopic parameters, such as the Coulomb interaction $V$, the electron-phonon interaction $g$, and the electronic hopping $J_0$. The general trend of a faster rise time with increasing fluence is surprising  due to the expected slowdown of the dynamics close to the critical  point~\cite{tsuji2013,golez2016,dolgirev2020,zong2019}. Before explaining this result, we contrast it to the dynamics of the order parameter.

As the order parameter is a complex number $\phi=|\phi| \exp[\I\theta]$, we will separately follow its absolute value $|\phi|$ and  phase $\theta$. Fig. 6 shows the time dependence of  $|\phi|$ and  $\theta$, as predicted by the model. After photo-excitation, $|\phi|$ diminishes with time until it levels off to a final, fluence-dependent value. For a large enough fluence, after a certain time, $|\phi|$ vanishes, which indicates the closing of the gap (dynamical critical point).  With increasing fluence, the overall time scale becomes longer on approaching the critical fluence, see Fig. 6(a). Intriguingly, the behaviour of the order parameter around the critical fluence, which was remarked also in different contexts~\cite{tsuji2013,golez2016,dolgirev2020,zong2019}, is in contrast with that manifested by the rise time of the electron population in the conduction band, see Fig.~\ref{Fig:tPES}(d).  This raises the following question: Can this slowing down also be observed in the photoemission spectra? It turns out that the momentum-integrated spectra do show such a behaviour. We extract the full width at
half maximum, $\sigma$, of the photoemission peak below the Fermi level, see Fig.~\ref{Fig:tPES}(e). The time dependence of this
quantity does become slower with fluence, see fitted rise times in Fig.~\ref{Fig:tPES}(f). The fact that $\sigma$ follows the dynamics of the order parameter can be understood by recalling that we are considering momentum integrated spectra: the suppression of the order parameter shows up in the modification of the dispersion that brings the upper edge of the valence band and lower edge of the conduction band closer; this leads to an increased width of the two~(momentum-integrated) bands.
Due to the limited energy resolution of our experiment, this theoretical prediction could not be directly verified with our present setup. We propose it as the subject of a future experimental study.  One could interpret this behaviour as a manifestation of the non-thermal criticality, as discussed in previous theoretical studies \cite{tsuji2013,golez2016, dolgirev2020}. 

\begin{figure}[t]
\includegraphics[width=1.0\linewidth]{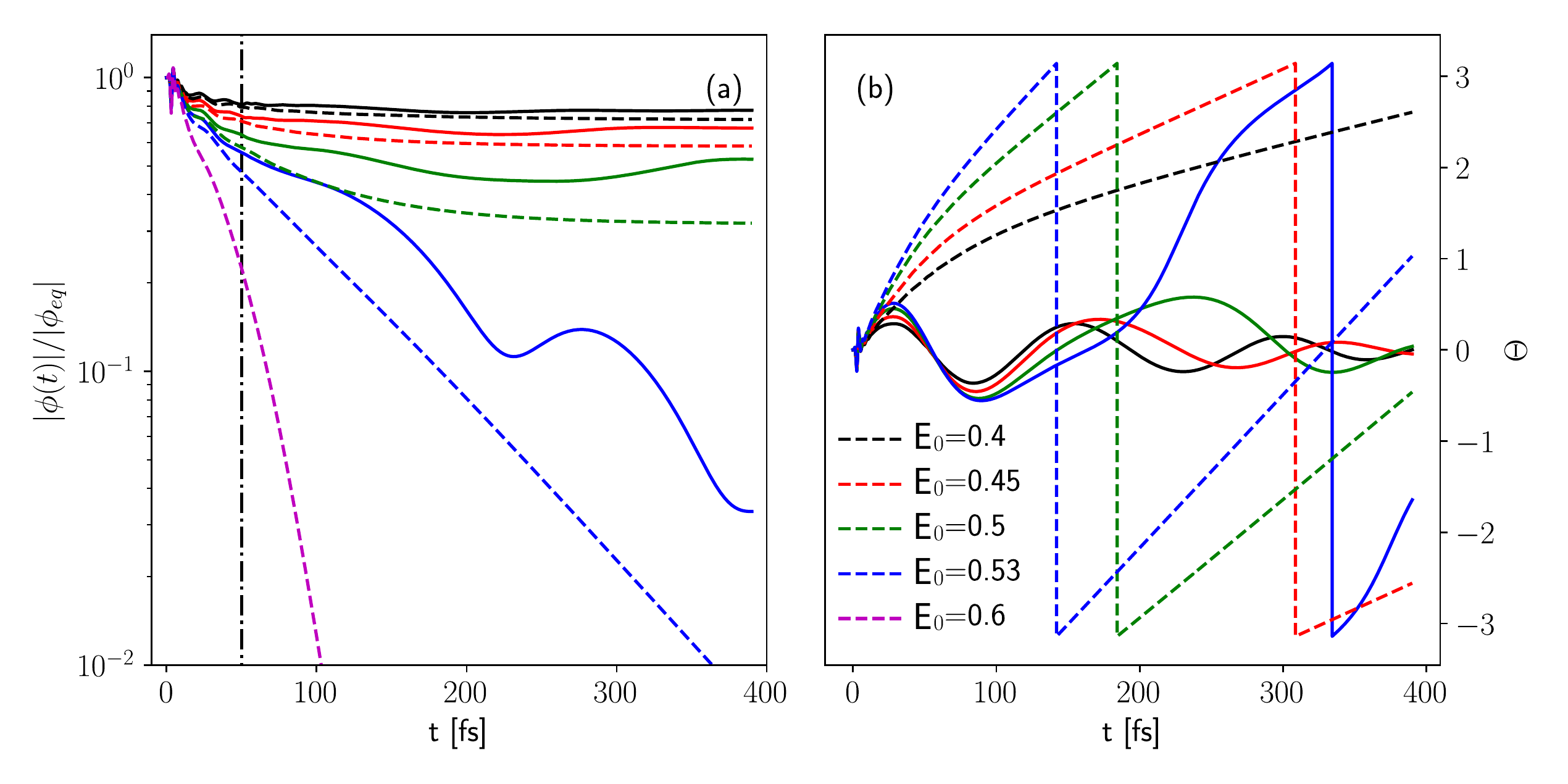}
\caption{Time evolution of the absolute value of the order parameter $|\phi|$~(a) and its phase $\theta$~(b), for different excitation strengths $E_0$. The dashed lines represent the evolution predicted by a purely electronic theory~($U/J_0$=3, $\lambda=0$), while the solid lines are the predictions including the electron-lattice coupling~($U/J_0$=2.94, $\lambda/J_0=0.03$). The vertical dashed-dotted line in a) separates the short and long time regions. For clarity, we have omitted the plot for $E_0=0.6$ in the purely electronic case in a).}
\label{Fig:order}
\end{figure}

While we have exposed the discrepancy between the behaviour of the signal's rise time and the order parameter vs. excitation strength, we would like to understand the origin of this dichotomy. If we look at early stages after the excitation ( $t<$ 50 fs) in Fig.~\ref{Fig:order}(a), the absolute value of $d |\phi(t)|/dt$ increases with  excitation strength, showing a speedup similar to the one observed for the rise time $\tilde\tau_r$~(there is no critical slowing down). In the later stages, this dynamics is altered and we observe critical slowing down of the order parameter with increasing fluence. However, we note that the absolute change of the order parameter in the early stages dominates over that in the later stage when the critical slowing down sets in. The short-time behaviour of the order parameter is consistent with a simple picture, where the condensed excitons act as a heat bath for relaxation of the photoexcited electrons. The energy transfer from electrons to the excitons leads to the decay of excitons and the process is faster if more electrons are excited; in this case, the order parameter also decays faster.

Based on the above considerations, we believe that the rise-time of the photoemission signal is related to the early-stages of the dynamics of the order parameter, where no critical slowing down is observed. The increase of the intensity in the momentum and energy-integrated spectrum can originate from two processes; i) relaxation of the excited electrons to lower energies due to scattering and ii) shift of the conduction band bottom into the integrated energy range. In i), if scattering of the high-energy electrons with condensed excitons is the dominant relaxation process, this relaxation stops when the excitons are completely consumed. Therefore, under strong photoexcitation, $|\phi|$ initially decays faster, and as the exciton bath is depleted, the relaxation also stops rapidly. In ii), because the order parameter is proportional to the band gap, the increase of the signal in the integrated region is directly proportional to the early-stages of the evolution of $|\phi|$. These two processes are therefore cooperating to shorten $\tilde\tau_r$ at high fluences.

We now discuss how the electron-phonon interaction modifies the above picture. In Fig.~\ref{Fig:order}, we compare the behaviour for the purely electronic case~($U/J_0$=3, $\lambda/J_0$=0) with the case including a weak electron-phonon interaction~($U/J_0$=2.94, $\lambda/J_0=0.03$). As a first remark, we note that a nonzero $\lambda$ value makes the order parameter $|\phi|$ more robust against photo-excitation, with the difference compared to the $\lambda=0$ case being more visible for larger excitation strengths, see Fig.~\ref{Fig:order}a.  
The importance of the electron-lattice interaction is even more dramatic for the phase of the order parameter $\theta$, see Fig.~\ref{Fig:order}b. The latter is either freely oscillating in the case $\lambda=0$, or confined in the case of electron-phonon coupling with~$\lambda/J_0=0.03.$ The deconfinement of the phase in the electron-lattice coupled situation for the strongest excitation, namely $E_0=0.53,$ coincides with an almost complete destruction of the order.  The robustness of the order parameter in the electron-phonon coupled systems can be understood as a combination of two effects: i) slower response of the lattice contribution to the gap closing; ii) opening of the finite mass of the phase mode of the excitonic order parameter, which prevents dephasing~\cite{murakami2017}. The latter effect only takes place for couplings that break the $U(1)$ gauge symmetry and make the phase mode massive~\cite{zenker2014,murakami2020_prb,guseinov1973,littlewood1996}. As the electron-phonon interaction is increased, $|\phi|$ becomes more robust after photo-excitation, and the phase more confined. 

Coming now to the effect of the electron-phonon interaction on the dynamics of the rise time $\tilde \tau_r$, shown in Fig.~\ref{Fig:tPES}(d), we considered the cases of different electron-phonon coupling strengths, namely $U/J_0$=2.94, $\lambda/J_0=0.03$ and $U/J_0$=2.8, $\lambda/J_0=0.1$.
The first observation is that the critical excitation strength increases with $\lambda$: $\Delta n=0.057$ for $\lambda/J_0=0.03$ and $\Delta n=0.07$ for $\lambda/J_0=0.1$, to be compared with $\Delta n=0.053$ for the case $\lambda=0$. This theoretical finding, together with the experimental observation that the gap closes only after a strong photo-excitation, indicates that electron-phonon coupling might play an important role in Ta$_2$NiSe$_5$, and provides a possible explanation for the observed small energy shifts in the trARPES spectra in Refs.~\onlinecite{mor2017,baldini2020} . Also in the cases with $\lambda \neq 0$, the rise  time shows a speed-up with increasing photo-excitation and the shape of the curve is similar to that found experimentally. Note however that in the curve corresponding to the largest $\lambda$ value, the relaxation time does not show any significant speedup with increased excitation strength below the critical one, in contrast to the experimental data. This gives an upper limit of the theoretical estimate of the electron-phonon coupling in the material. Similar conclusions are obtained for the time scale of the broadening $\tau_{\sigma},$ see Fig.~\ref{Fig:tPES}(f). The main signature of the increased electron-phonon coupling is an increase of the critical excitation strength.

\subsection{Phenomenological description}
As the microscopic calculations presented in previous sections are rather involved, we will introduce a phenomenological model, which provides a simple explanation of the initial speedup of order parameter dynamics with increasing fluence, followed by a slowdown when approaching the critical excitation strength. The basic assumption of the model is that the condensed excitons act as a heat bath for photoexcited electrons. We have  confirmed this assumption by explicit calculations of photo-excitation in the semi-metallic and semiconducting states without excitonic order and found that the relaxation is much weaker and slower as compared to the relaxation in the case of an excitonic insulator~(data not shown).

The model considers, besides the order parameter $\phi$ (for brevity, we will use the notation $\phi$ instead of $|\phi|$ in the following), also excited electrons that are characterized by a density $\mathcal{N}$. The relaxation of  electrons is assumed to dominantly take place through coupling with excitons and vice-versa - the relaxation of the excitonic order parameter occurs through coupling with $\mathcal{N}$.  The probability of the relaxation process is proportional to the number of excitations and to the number of excitons, i.e. to $\mathcal{N}$ and $\phi$. Hence, the two equations that govern the behaviour of the two coupled variables can be written as 
\begin{equation}
  d \phi /dt =  -c_\phi \phi \mathcal{N}   , 
  \end{equation}
\begin{equation}
  d \mathcal{N}/dt =  -c_\mathcal{N} \phi \mathcal{N},
\end{equation}
where the $c_{\phi}$ is the exciton dissociation rate due to scattering with high-energy electrons and  $c_\mathcal{N}$ is  the decay rate of excited electrons. Let us first discuss the extreme cases of very weak and very high fluences. In the former case, the value of the order parameter remains close to the initial one, $\phi_I$. Conversely, $\mathcal{N}$ is exponentially damped  and its relaxation time 
is given by $1/(c_\phi \phi_I)$, i.e., it becomes short for large values of the order parameter $\phi$. In the latter case, $\mathcal{N}$ and $\phi$ exchange their roles and now $\phi$ is exponentially damped  and its relaxation time is given by $1/(c_\mathcal{N} \mathcal{N}_I)$, i.e., it becomes short for large values of excitation $\mathcal{N}$. 

Away from those two extremal regimes, one can integrate the two 
equations. Because the two time derivatives are proportional to each other, $d \mathcal{N}/dt = (c_\mathcal{N}/c_\mathcal{\phi}) d \phi/dt$ (in this approximation, the relaxation of an electron implies dissociation of an exciton), $\phi$ and $\mathcal{N}$ are linearly related
\begin{equation}
  \mathcal{\phi} = c_\phi /c_\mathcal{N} \mathcal{N}  + \mathrm{const.}
\end{equation}
This enables rewriting of the problem into a simple differential equation that can be integrated. For weak fluences (below the critical one), the long-time value of $\mathcal{N}$ vanishes, but the order parameter reaches a nonvanishing value $\phi_F$, hence one takes $\mathcal{N} = (\phi-\phi_F) c_\mathcal{N}/c_\phi $.
This leads to a solution~(denoting the initial condition as $\phi(t=0)=\phi_I$)
\begin{equation}
  \phi(t) = \frac{\phi_I \phi_F e^{t / \tau_\phi}}{\phi_F-\phi_I+\phi_I e^{  t / \tau_\phi}}, \label{eq:phi}
\end{equation}
where $1/\tau_\phi = c_\mathcal{N} \phi_F$.
At long times, $\phi$ thus decays towards $\phi_F$ with a decay time proportional to $1/\phi_F$, which directly implies a critical slowing down on approaching the critical fluence. On the other hand, at short times, the decay is faster than that inferred from continuation of exponential decay to short times. More precisely, the initial slope $d \phi/dt = -c_\mathcal{N} \phi_I \Delta \phi$ is proportional to the change of the order parameter $\Delta \phi=(\phi_I-\phi_F)$; i.e., the speed of the initial decay is higher for increasing fluence, which explains the short-time behaviour of the order parameter and the rise time in numerical simulations.

\begin{figure}[t]
  \includegraphics[width=1.0\linewidth]{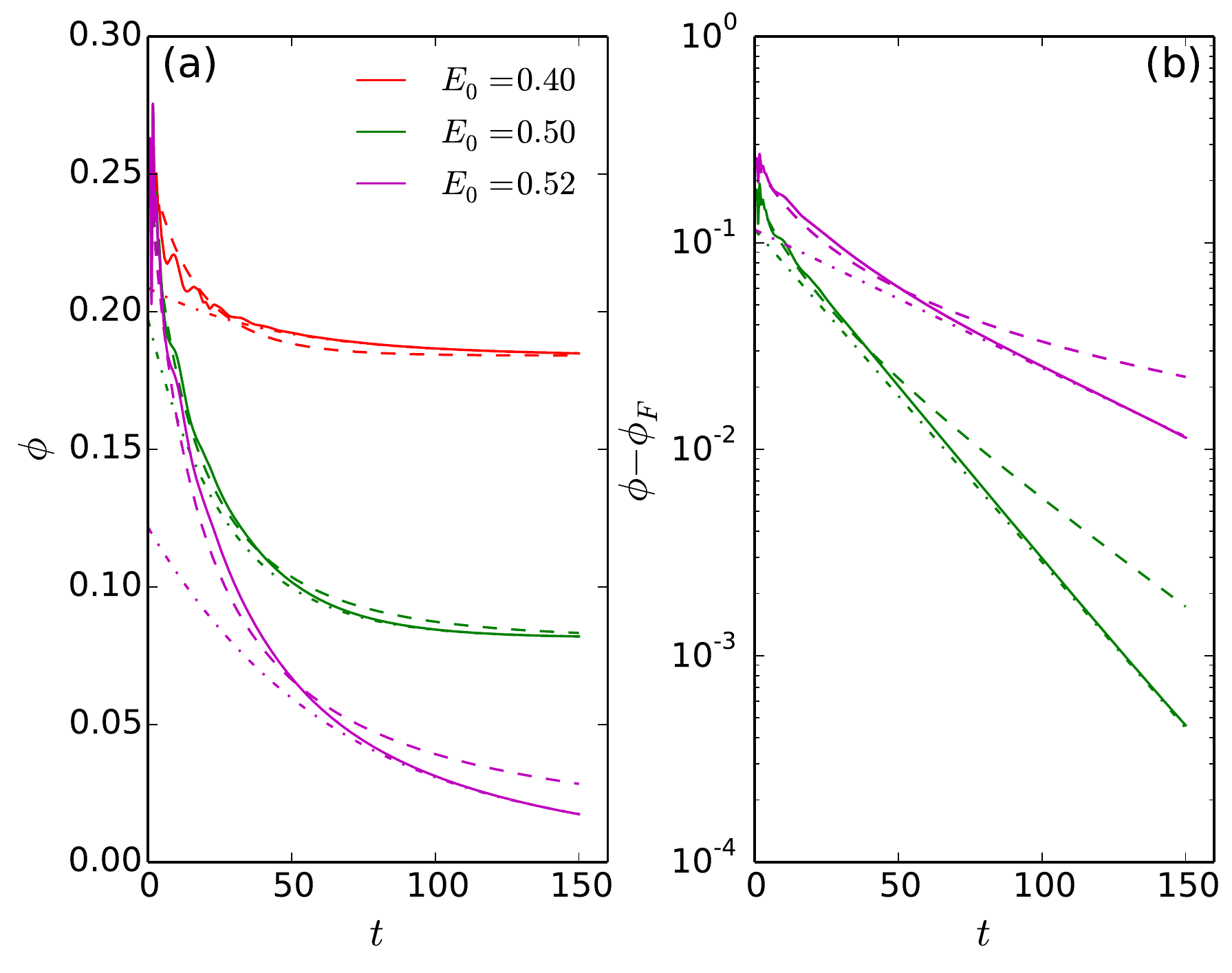} \\
  \includegraphics[width=1.0\linewidth]{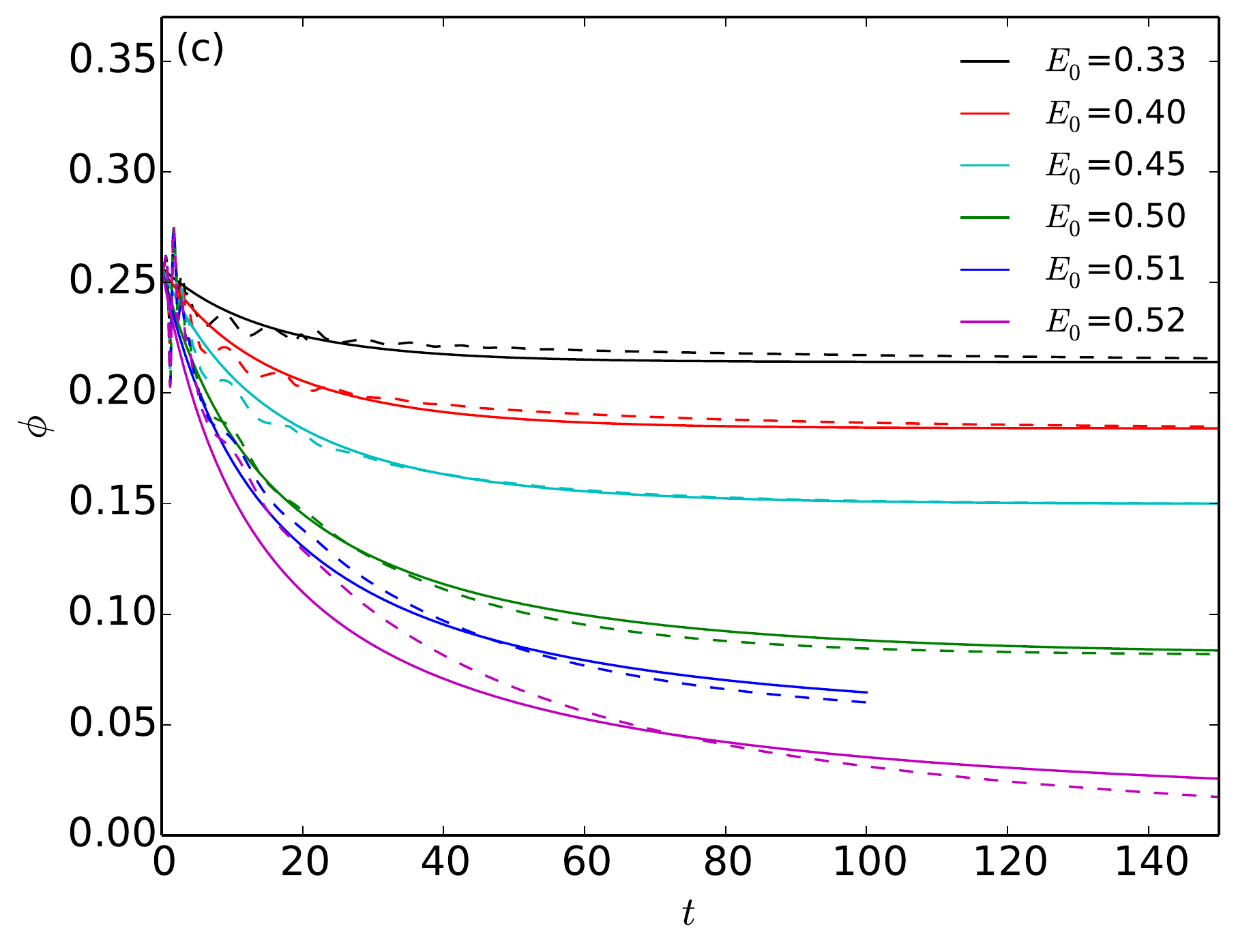}
\caption{(a) Absolute value of the order parameter $\phi$ (full) fitted with a heuristic function (dashed) and compared also to an exponential decay towards the long-time value $\phi_F$ (dashed-dotted). (b) Logarithmic plot of $\phi-\phi_F$. (c) Parameterless fits to data. } 
\label{Fig:heur1}
\end{figure}

The fit of Eq.~\ref{eq:phi} to the numerical data is shown in Fig. ~\ref{Fig:heur1}(a). One sees that at short times $\phi$ indeed decays faster, consistent
with the heuristic function, and yet the slope of the approach towards  $\phi_F$ at long-times, shown in Fig. 7(b), becomes smaller; the order parameter dynamics slows down as the long-time value $\phi_F$ becomes small (note that the the slope is  reproduced only qualitatively).   Fitted $c_\mathcal{N}$ depend on each case very weakly and a single value can be used for all fluences. This is demonstrated in Fig. \ref{Fig:heur1}(c)  where we took $c_\mathcal{N}$=3.7 for all the fluences. Note that $\phi_F$ is independently determined from the
long-time value, and there are thus no free parameters.

Finally, it is interesting to consider also fluences above the critical one, where $\phi_F=0$ and there is a remaining density of excitations $\mathcal{N}_F$ in the limit of long times. In this case $\phi$ decays with a rate $1/\tau_N \propto \mathcal{N}_F$, that is, the decay becomes faster at higher fluences, which is indeed what one sees numerically, cf. Fig.~\ref{Fig:order}. Moreover, one can estimate $\mathcal{N}_F$ for fluences above the critical one from the behaviour for fluences below the critical one and find that the long-time tails indeed decay proportionally to $\mathcal{N}_F$. 

The key observations of the phenomenological analysis are thus: (i) the critical slowing down of the dynamics of $\phi$ occurs on approaching the critical fluence  and is described naturally by the lack of  excitons acting as an efficient bath, (ii) this slowing down of the dynamics occurs in the limit of long times only and is related to the fact that the time scale is given by the inverse value of the order parameter (which itself becomes small), (iii) the dominant change of the order parameters happens at short times, where the derivative is independent of time for fluences below the critical one, (iv) for fluences above the critical one, the evolution becomes faster also initially.

\section{Conclusion}
Summing up, our experimental results demonstrate that a phase transition in Ta$_2$NiSe$_5$ from an insulating to a semi-metallic state can be induced under strong photoexcitation, which belongs to a fluence regime that has not been explored before\citep{mor2018,okazaki2018}. We observe that the closing of the band gap occurs on a time scale faster than the typical phonon oscillation in the material. Furthermore, we find that the (fast) rise time of photo-emission intensity above the Fermi energy and the (slow) relaxation time of the electron population towards equilibrium show contrasting behaviours as a function of the excitation strength: while the rise time becomes shorter with increasing fluence, the relaxation time towards the equilibrium becomes longer, reaching saturation above the critical fluence.

The observed speed up of the photoemission signal is confirmed by the results provided by a theoretical two-band tight-binding model, which allowed us to interpret the experimental trends in terms of scattering of the photoexcited electrons with the excitonic condensate. We found that the inclusion of a phonon mode does not modify the physics qualitatively, however, it makes the model system more robust towards photoexcitation. This might explain the apparent discrepancies between different trARPES studies reported in the literature \cite{mor2017,okazaki2018,baldini2020,tang2020,suzuki2020detecting}. We note, however, that this aspect needs to be considered in the future in a more realistic setting with genuine symmetry of the involved phonon mode and the excitonic order parameter~\cite{subedi2020_prm}. 

Finally, the above results point to a predominately electronic origin of the order parameter in Ta$_2$NiSe$_5$ and allow us to develop simple phenomenological model, which describes the dynamics after photoexcitation in an excitonic insulator.

\acknowledgements

D. G. and J. M. acknowledge the support of the Slovenian Research Agency (ARRS) under Programs No. J1-2455, J1-1696, J1-2458, and P1-0044. J. M. acknowledges insightful discussions with Alaska Subedi. The Flatiron Institute is a division of the Simons Foundation. Y.M. acknowledges the support by a Grant-in-Aid for Scientific Research from JSPS, KAKENHI Grant Nos. JP19K23425, JP20K14412, JP20H05265, and JST CREST Grant No. JPMJCR1901. This work has received funding from the European Union’s Horizon 2020 research and innovation programme under grant agreement No 654360 NFFA-Europe. We thank A. Ciavardini for valuable help during experiments.

\appendix
\section{Application of the time-dependent second Born approximation to the excitonic insulator}\label{App:2Born}

To allow for the excitonic symmetry breaking, it is useful to introduce
the spinor \beq{ \Psi_{k,\alpha}\equiv
\begin{pmatrix}
c_{k,0}  \\
c_{k,1}  \\
\end{pmatrix}
} and the corresponding 2$\times$2 Green's function is given by \beq{
  G_k(t,t')=-\I \green{\Psi_k}{\Psi^\dagger_k} } and it is determined from a
solution of the Dyson's equation. For latter usage we will introduce
the single-particle density matrix
$\rho_{k,\alpha\beta}=\ave{\Psi^{\dagger}_{k,\beta} \Psi_{k,\alpha}},$
where the Greek indices indicate the orbital space. The local
component of the density matrix is given by
$\rho_{\text{loc},\alpha\beta}=\sum_k \rho_{k,\alpha\beta}.$ The
occupation of the valence~(conducting) band is therefore given by
$n_0=\rho_{\text{loc},00}~(n_1=\rho_{\text{loc},11}).$ Similarly, the
excitonic order parameter is given by the off-diagonal component of
the density matrix $\phi=\rho_{\text{loc},01}.$ The details of the
theoretical solution are presented in the following.

The employed approximation is determined by the self-energy and we will treat the Coulomb interaction within the 2nd Born approximation, which includes the inelastic scattering and captures the  heating effects.  For the Coulomb interaction, the Hartree term is given by
\beq{
	\Sigma^H_{k,ij}(t)=\delta_{ij} V \rho_{\text{loc},\bar i\bar i}(t),
}
where the overline marks the oposite band and the Fock term is given by
\beq{
	\Sigma^F_{k,ij}(t)= V\rho_{\text{loc},ij}(t).
}
The coupling with phonons will be described on the Hartree-Fock level with the instantaneous term 
\beq{
	\Sigma^{ph}_{k,ij}(t)= g X(t) \delta_{|i-j|=1},
}
and the phonon motion is described by a system of differential equations for the phononic distortion $X$ and the corresponding momentum $P$
\bsplit{
	&\partial_t P(t)=\omega_0 P(t) \\  
	&\partial_t X(t)=-\omega_0 X(t) -2 g (\rho_{\text{loc},01}+\rho_{\text{loc},10}).
}

\begin{figure}
\includegraphics[width=\linewidth]{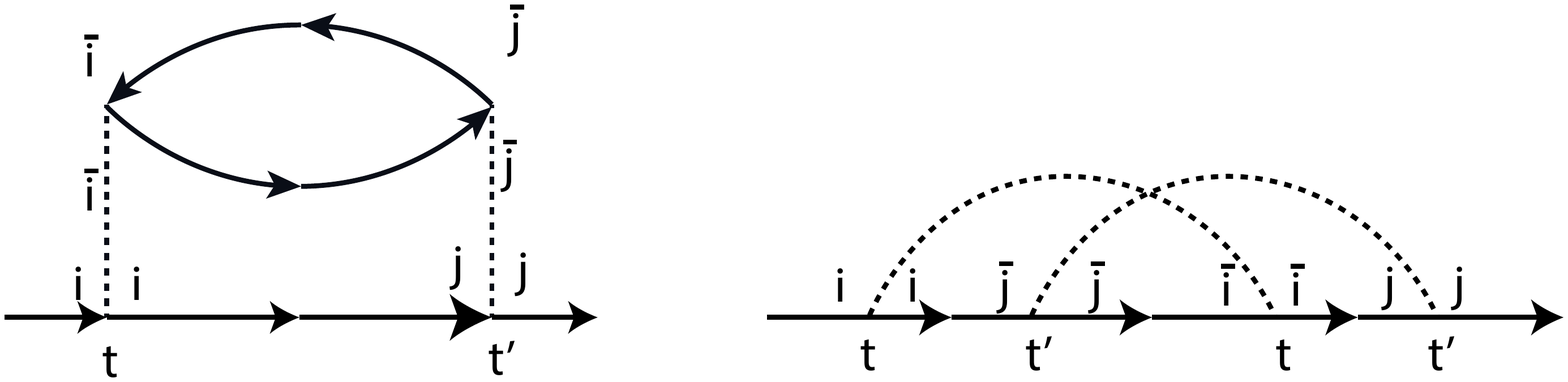}\\
\caption{The interacting self-energy diagrams corresponding to the 2nd Born approximation.}
\label{Fig:Selfenergy}
\end{figure}

As the Hartree and Fock self-energies are instantaneous, it is useful to redefine the single particle Hamiltonian as $h_k^{HF}=h_k+\Sigma^H_k+\Sigma_k^F$ and introduce the renormalized free propagator as
\beq{
	G_k^{HF}=(\I\partial_t-h_k-\Sigma^H-\Sigma^F_k-\Sigma^{ph}_{k})^{-1}
}

In the 2nd Born approximation, the correlation self-energy is given by the expansion in the interaction strength which is then resumed using the Dyson equation, see Fig.~\ref{Fig:Selfenergy} for the diagrammatic representation of the self-energy. The analytical expression for the self-energy is given by
\bsplit{
\Sigma^{2B}_{q,ij}(t,t')=&-\I V^2 \sum_k  G_{k,\bar j\bar i}(t',t) \\
&\left [\chi_{k+q,i\bar j}(t,t')- \chi_{k+q,\bar i\bar j}(t,t')\right],}
\label{Eq:Sigma}
where we have introduced the charge susceptibility bubble
\beq{ 
\chi_{q,ij}(t,t')=\I \sum_k G_{q-k,i\bar j}(t,t')G_{k,\bar i j}(t,t’). 
\label{Eq:Pol}
}

The Greens function is then obtained from the solution of the Dyson equation
\beq{
	G_k=G_k^{HF} + G_k^{HF}\ast\Sigma_k^{2B}\ast G_k,
}
where $\ast$ denotes the convolution in time and the matrix multiplication in the orbital space. We solve this problem numerically using the library NESSi employing an efficient MPI parallelization over the momentum points~\cite{schuler2020cp}.

\end{document}